\documentclass[prb,preprint]{revtex4}
\usepackage{graphicx}% Include figure files
\usepackage{dcolumn}% Align table columns on decimal point
\usepackage{bm}% bold math
\usepackage{amssymb,amsmath}

\begin{document}
\title{Linking high and low temperature plasticity in bulk metallic glasses II: use of a log-normal barrier energy distribution and a mean field description of high temperature plasticity}
\author{P. M. Derlet}
\email{Peter.Derlet@psi.ch} 
\affiliation{Condensed Matter Theory Group, Paul Scherrer Institut, CH-5232 Villigen PSI, Switzerland}
\author{R. Maa{\ss}}
\email{robert.maass@ingenieur.de}
\affiliation{Institute f\"{u}r Materialphysik, University of G\"{o}ttingen, Friedrich-Hund-Platz 1, D-37077 G\"{o}ttingen, Germany}
\date{\today}

\begin{abstract}
A thermal activation model to describe the plasticity of bulk metallic glasses (Derlet and Maa\ss, Phil. Mag. 2013, DOI: 10.1080/14786435.2013.826396) which uses a distribution of barrier energies and some aspects of under-cooled liquid physics is developed further. In particular, a log-normal distribution is now employed to describe the statistics of barrier energies. A high temperature mean-field description of homogeneous macro-plasticity is then developed and is shown to be similar to a thermal activation picture employing a single characteristic activation energy and activation volume. In making this comparison, the activation volume is interpreted as being proportional to the average mean-square-value of the plastic shear strain magnitude within the material. Also, the kinetic fragility at the glass transition temperature is shown to represent the effective number of irreversible structural transformations available at that temperature.
\end{abstract}
\maketitle

\section{Introduction}

The deformation properties of bulk metallic glasses are characterised by two temperature regimes~\cite{Schuh2007}. At high temperatures, close to the glass transition temperature, BMGs deform homogeneously with the strain rate properties being reasonably well understood by thermally activated plastic flow. At lower temperatures, plasticity becomes heterogeneous and highly localized to shear-bands, resulting in limited ductility in tension and a few percent plastic strain prior to failure for some BMGs in compression. In this regime of temperatures both athermal~\cite{Falk2011} and thermal~\cite{Spaepen1977,Argon1979,Johnson2005} theories of plasticity have been proposed. In the early work of Turnbull and Cohen~\cite{Cohen1959,Turnbull1961,Turnbull1970} these different regimes emerged as a result of two competing processes, one diffusional and one stress induced. Later, Spaepen \cite{Spaepen1977} attributed the homogeneous regime as resulting from a steady state of free volume creation and diffusional related free volume annihilation. Heterogeneous flow on the other hand was dominated by free volume creation since annihilation became diffusion limited at low temperatures. That two distinct temperature scales exist was also recognised by Argon~\cite{Argon1979}, who from the perspective of thermal activation developed two microscopic theories where at high temperature, plasticity was mediated by small atomic reconfigurations that resulted in a localized shear strain event, and at lower temperature, a less local structural excitation that was akin to the nucleation of a dislocation loop occurs. Such mechanisms could give insight into the observed high temperature homogeneous plasticity and the low temperature heterogeneous plasticity.

Some time after this, high-strain rate atomistic simulation revealed the generality of such local structural excitations, where the transition from homogeneous to heterogeneous plasticity (as a function of decreasing temperature or increasing strain rate) emerged from the degree of correlation between such localized activity~\cite{Falk1998,Schuh2003,Maloney2004,Demkowicz2004,Shi2006,Guan2010}. Due to their high strain rates and stresses, both dynamic and static atomistic simulations primarily probe the athermal region of plasticity. Out of the earliest of this work by Falk and Langer~\cite{Falk1998} arose the concept of the Shear Transformation Zone (STZ) which was assumed to exist in regions of the material not so well relaxed --- the so-called liquid like regions \cite{Demkowicz2004,Demkowicz2005,Demkowicz2005a,Langer2008}. The corresponding structural transformations were assumed to be activated {\em athermally} once a critical stress had been reached. Such ideas developed into what is now called the effective temperature theories in which the quenched disorder well below the glass transition temperature is characterised by an effective temperature that describes the structural fluctuations within the material, see for example~\cite{Falk2011}. This approach has been quite successful in developing a quantitative understanding of experimental low temperature deformation data.

It was recognised by Bouchbinder, Langer and Falk~\cite{Bouchbinder2007} that if one were to assume a thermal activation picture within the STZ framework~\cite{Falk2004,Langer2004}, the density of such local structural excitations would have to be ``improbably'' large if the corresponding activation energies are in the realistic range of hundreds of kilojoule per mole (or electron volts per excitation). In going to an athermal picture, this difficulty was avoided and indeed STZs could now occur at quite dilute densities~\cite{Bouchbinder2007}, a picture that is compatible with the distinct liquid like regions assumed to exist within the glass.

Understanding the nature of such STZs, or more generally localized structural excitations, has also been influenced by the physics of the under-cooled liquid regime, in which two distinct relaxation time-scales are evidenced: that of the slow $\alpha$-relaxation and the fast $\beta$-relaxation processes first introduced by Goldstein~\cite{Goldstein1969}. At temperatures near the glass transition and below, the $\alpha$-relaxation processes rapidly increase their time-scale with the traditional viewpoint being that they freeze out of the structural dynamics in the amorphous solid regime~\cite{Angell2000}. Despite such an assumed freezing, the terminology of $\alpha$- and $\beta$-relaxation has also been applied to the amorphous solid regime where the microscopic $\beta$ processes mediate escape from the mega-basins of the elastic energy landscape of the $\alpha$ processes~\cite{Harmon2007,Wang2011}. The present work also exploits this terminology and uses an additional fundamental feature of the under-cooled liquid regime, that the number of structural transformations available will scale exponentially with respect to a fundamental volume measure~\cite{Shell2004,Fyodorov2004}. Such ideas have been used in the early thermodynamic theories of Adam and Gibbs~\cite{Adam1965} and Kirkpatrick {\em et al} \cite{Kirkpatrick1989} --- see also the review article by Heuer~\cite{Heuer2008} --- and provides a sufficient number of structural transformations to justify a thermal activation picture for temperatures that are well below the glass transition.

In particular the present work develops further the ideas of refs.~\cite{Derlet2011,Derlet2012,Derlet2013}. In these works, a thermal activation picture was developed in which the characteristic plastic rate with respect to an internal heterogeneous volume scale was determined from a distribution of $\alpha$-relaxation barriers whose number scales exponentially with the size of the heterogeneous volume and whose statistics is described by a distribution with extensive first and second cumulants. By doing this, the average time scale associated with plasticity at a particular temperature becomes intimately connected to the $\alpha$-relaxation potential energy landscape (mediated by $\beta$ relaxation processes) which is thermally accessible at that temperature. At zero loading or any given shear stress, two critical temperature scales emerge: 1) the plastic transition temperature which indicates a significant drop in the amount of plastic activity as the temperature is reduced, and 2) the (lower) kinetic freezing temperature where (on average) a single structural transformation becomes available per heterogeneous volume. The transition between the higher temperature regime of 1) and the lower temperature region of 2) corresponds to a change of statistics, from that of the most probable to that of the extreme. Importantly, before the critical freezing temperature is reached, there will exist a low temperature regime (below the glass transition) where a sufficient number of thermally activated events exist to mediate a non-negligible macroscopic plasticity. Thus, in ref.~\cite{Derlet2013}, the low temperature heterogeneous and high temperature homogeneous deformation regimes are distinguished by a transition in the nature of the statistics of the $\alpha$-relaxation energy landscape. Put in other words, the freezing out of $\alpha$-relaxation processes is only complete at temperatures far below that of the glass transition temperature, and until this extreme regime is reached thermal activation remains the underlying phenomenon controlling macroscopic plasticity --- a fact that has been established experimentally for a variety of deformation geometries~\cite{Schuh2007} including most recently that of shear band nucleation, propagation and arrest~\cite{Klaumunzer2010,Maass2011,Maass2012}.

The present work extends on the work done in ref.~\cite{Derlet2013} (which will be referred to as paper I) by using a log-normal distribution to describe the statistics of the $\alpha$-relaxation barrier energy landscape to develop a mean-field description of thermally activated plasticity for the high temperature/low strain rate regime of homogeneous deformation. Sec.~\ref{SecSummary} gives a brief overview of the current theory, secs.~\ref{SecLND}-\ref{SecTheory} develops the theory for the positive valued log-normal distribution of barrier energies, sec.~\ref{SecMFP} develops a mean field description of plasticity suitable for the high temperature homogeneous regime of deformation, and sec.~\ref{SecExp} applies the theory to well known experimental data for the well studied BMG Vitroley-1. In the discussion and conclusion, sec.~\ref{SecConc}, two results are summarized: 1) when comparing with the usual Arrhenius description of high temperature homogeneous plasticity, the corresponding activation volume parameter is found to be equivalent to the variance of the internal slipped volume distribution, and 2) the fragility of the material is a measure of the number of structural transformations available at the glass transition temperature.

\section{Brief summary of basic theory} \label{SecSummary}

Paper I considered a characteristic time scale of irreversible structural transformation activity, $\tau_{\mathrm{p}}$, whose inverse is seen as a plastic transition rate below the glass transition temperature. Studying the resulting temperature and stress dependence of $\left[\tau_{\mathrm{p}}\right]^{-1}$ gave insight into the transition from elasticity to micro-plasticity. To determine $\left[\tau_{\mathrm{p}}\right]^{-1}$ it was recognised that, at a large enough length scale, BMGs are considered structurally homogeneous indicating that a sufficient amount of self-averaging occurs with respect to a shorter heterogeneous length scale. Thus two length-scales naturally emerge which are realised through the homogeneous and heterogeneous volume elements. The plastic rate for a homogeneous representative volume element (RVE) of volume $V_{\mathrm{RVE}}$ is
\begin{equation}
\left[\tau_{\mathrm{RVE,p}}\right]^{-1}(T)=\sum_{n=1}^{N'}\left[\tau_{\mathrm{p,n}}\right]^{-1}(T)=N'\times\left[\tau_{\mathrm{p}}\right]^{-1}(T)
\end{equation}
in which 
\begin{equation}
\left[\tau_{\mathrm{p}}\right]^{-1}(T)=\frac{1}{N'}\sum_{n=1}^{N'}\left[\tau_{\mathrm{p},n}\right]^{-1}(T). \label{EqnTauP}
\end{equation}
In the above $\left[\tau_{\mathrm{p},n}\right]^{-1}$ is the particular plastic rate for the $n$th heterogeneous volume element and $N'$ is the number of such heterogeneous volume elements leading to a well-converged self-averaged $\left[\tau_{\mathrm{p}}\right]^{-1}$. For the current work, the characteristic volume of the heterogeneous volume is labelled as $V_{0}$ giving $N'=V_{\mathrm{RVE}}/V_{0}$.

Under the assumption of thermally activated plasticity, the plastic rate associated with one particular heterogeneity volume is written as a linear sum of the $M=M(N)$ thermally active transition rates available to that volume element:
\begin{equation}
\left[\tau_{\mathrm{p,n}}\right]^{-1}(T)=\sum_{i=1}^{M}\left[\tau_{\mathrm{p0},ni}\right]^{-1}(T)\exp\left(-\frac{E_{\mathrm{p0},ni}}{k_{\mathrm{B}}T}\right), \label{EqnTauPn}
\end{equation}
where $\left[\tau_{\mathrm{p0},ni}\right]^{-1}(T)$ and $E_{\mathrm{p0},ni}$ are the attempt rate and barrier energy for the $i$th irreversible structural transformation within the $n$th heterogeneous volume element. Eqn.~\ref{EqnTauP} then becomes
\begin{equation}
\left[\tau_{\mathrm{p}}\right]^{-1}(T)=\left\langle\left[\tau_{\mathrm{p0}}\right]^{-1}(T)\exp\left(-\frac{E_{\mathrm{p0}}}{k_{\mathrm{B}}T}\right)\right\rangle \label{EqnAveTauP}
\end{equation}
in which the average is performed with respect to $\left[\tau_{\mathrm{p0}}\right]^{-1}(T)$ and $E_{\mathrm{p0}}$.

In terms of the underlying potential energy landscape (PEL), the simple (first order) expression of eqn.~\ref{EqnTauPn} is only valid for thermally activated processes that do not multiply recross their energy barrier. This naturally leads to a coarse graining of the PEL, in which the barriers and (now diffusive) pre-factors entering into eqn.~\ref{EqnTauPn} underlie the collective microscopic activity that results in a lasting escape from a characteristic energy valley. By analogy to the under-cooled liquid PEL framework~\cite{Heuer2008,Harmon2007} where two distinct time scales occur --- the ``slow'' $\alpha$-relaxation and ``fast'' microscopic $\beta$-relaxation modes --- the self-averaging of eqn.~\ref{EqnAveTauP} reduces (see paper I) to
\begin{equation}
\left[\tau_{\mathrm{p}}\right]^{-1}=\left[\tau_{\mathrm{p00}}\right]^{-1}\exp\left(-\frac{E_{\mathrm{p00}}}{k_{\mathrm{B}}T}\right)\times M\left\langle\exp\left(-\frac{E}{k_{\mathrm{B}}T}\right)\right\rangle. \label{EqnPlasticRate}
\end{equation}
Eqn.~\ref{EqnPlasticRate} consists of a diffusive attempt rate with a simple Arrhenius temperature dependence, representing the mediating $\beta$ mode relaxation dynamics multiplied by a thermal factor whose temperature dependence will be derived from the statistical properties of the $\alpha$-relaxation mode coarse grained PEL~\cite{Derlet2012}.

In paper I the number of structural transformations available to each heterogeneous volume was assumed to scale exponentially with the number of atoms within $V_{0}$, giving $M=\exp(\alpha)$ where $\alpha=\overline{\alpha}N$ and $N$ is the characteristic number of atoms within a heterogeneous volume. The right most average in eqn.~\ref{EqnPlasticRate} is then evaluated via a distribution of barrier energies defined by an extensive first and second cumulant, and written as
\begin{equation}
M\left\langle\exp\left(-\frac{E}{k_{\mathrm{B}}T}\right)\right\rangle=\exp\left[-\frac{E(T)-TS(T)}{k_{\mathrm{B}}T}\right]=\exp\left[-\frac{F(T)}{k_{\mathrm{B}}T}\right]. \label{EqnFE}
\end{equation}
Here $E(T)$ is interpreted as the temperature dependent internal barrier energy and $S(T)$ as the temperature dependent barrier entropy. Within this context, $F(T)$ is referred to as the free barrier energy. All of these quantities are extensive and $\exp(S(T)/k_{\mathrm{b}})$ gives the apparent number of structural transformations available at temperature $T$. In paper I, a Gaussian distribution was used giving closed form expressions for both $E(T)$ and $S(T)$. An important property of the above is that for a large enough $N$ there exists a sharp crossover in behaviour where for, $F(T)<0$, eqn.~\ref{EqnFE} is negligible and for, $F(T)>0$, it is exponentially large. Thus $T_{\mathrm{c}}$, where $F(T_{\mathrm{c}})=0$, defines a critical temperature at which the plastic transition rate rapidly rises.

By assuming the application of a pure shear stress broadens the distribution via a quadratic increase in the variance (with respect to stress), all mentioned quantities, including $T_{\mathrm{c}}$, gain a shear stress dependence. Via the assumption that at zero applied shear stress $T_{\mathrm{c}}$ corresponds approximately to the glass transition temperature, the stress dependence of $T_{\mathrm{c}}$ rises rapidly from zero at $T_{\mathrm{g}}$ as the temperature is lowered. At a temperature of approximately 0.8-0.9$T_{\mathrm{g}}$ this temperature dependence weakens to an approximately linear behaviour upon further reduction in temperature. Interpreting $T_{\mathrm{c}}(\sigma_{\mathrm{c}})=T_{\mathrm{Deformation}}$ (where $T_{\mathrm{Deformation}}$ is the temperature at which the deformation experiment is performed) as being the stress at which plasticity begins, an analytical form of yield stress versus temperature that is valid in both the low and high temperature regimes could be developed, and could be fitted well to the available experimental data (see fig.~8 of \cite{Derlet2013}) of Lu {\em et al} \cite{Lu2003} and that shown in Johnson and Samwer~\cite{Johnson2005}. Whilst a successful fit to low temperature yield data is not new, the current model has the virtue that the yield stress rapidly decreases as the temperature approaches $T_{\mathrm{g}}$ due to a change in barrier energy statistics rather than a change in mechanism. 

\section{The log-normal barrier energy distribution} \label{SecLND}

In paper I a normal distribution of barrier energies was considered and also a modified version such that the distribution would limit to zero at zero barrier energy. The present work will consider a log-normal distribution of barrier energies. This distribution has the desirable features of being a positive valued distribution with independent first and second cumulants (average and variance) --- a requirement of the current model. The log-normal distribution is given as
\begin{equation}
P(E)=\frac{1}{\sqrt{2\pi}E\sigma}\exp\left[-\frac{\left(\log E-\mu\right)^{2}}{2\sigma^{2}}\right] \label{EqLND0}
\end{equation}
where the first moment, $\left\langle E\right\rangle$, is 
\begin{equation}
\left\langle E\right\rangle=\exp\left[\mu+\frac{1}{2}\sigma^{2}\right]
\end{equation}
and the second moment, $\left\langle E^{2}\right\rangle$, is defined via
\begin{equation}
\sigma^{2}=\log\left[1+\frac{\left\langle E^{2}\right\rangle-\left\langle E\right\rangle^{2}}{\left\langle E\right\rangle^{2}}\right].
\end{equation}
In terms of $\left\langle E\right\rangle$ and $\left\langle E^{2}\right\rangle$, the distribution can be written as
\begin{equation}
P(E)=\frac{1}{\sqrt{2\pi}E\sqrt{2\log a}}\exp\left[-\frac{\log^{2}\left[\frac{Ea}{\left\langle E\right\rangle}\right]}{4\log a}\right] \label{EqLND}
\end{equation}
with
\begin{equation}
a=\frac{\sqrt{\left\langle E^{2}\right\rangle}}{\left\langle E\right\rangle}.
\end{equation}

In paper I it was argued that the barrier energy distribution shall have extensive first and second cumulants, that is, the barrier energy mean $E_{0}$ and variance $\delta E_{0}^{2}$ will be linear functions of the number of atoms, $N$, within the heterogeneous volume element. As in paper I, an over-lined quantity will represent the corresponding intensive variable giving $E_{0}=N\overline{E}_{0}$ and $\delta E_{0}^{2}=N\delta\overline{E}_{0}^{2}$. Thus, the first and second moments of eqn.~\ref{EqLND} will be
\begin{eqnarray}
\left\langle E\right\rangle&=&E_{0}=N\overline{E}_{0} \label{EqnE0} \\
\left\langle E^{2}\right\rangle&=&E_{0}^{2}+\delta E_{0}^{2}=N^{2}\overline{E}_{0}^{2}\left(1+\frac{1}{N}\left(\frac{\delta\overline{E}_{0}}{\overline{E}_{0}}\right)^{2}\right) \label{EqnE20}
\end{eqnarray}
resulting in
\begin{equation}
a=\sqrt{1+\left(\frac{\delta E_{0}}{E_{0}}\right)^{2}}=\sqrt{1+\frac{1}{N}\left(\frac{\delta\overline{E}_{0}}{\overline{E}_{0}}\right)^{2}}. \label{Eqna}
\end{equation}

The log-normal distribution in the form of eqn.~\ref{EqLND} will now be used to obtain the average plastic rate per heterogeneous volume (sec.~\ref{SecTheory}) and, through a mean-field description of the internal stress, the average plastic strain rate per heterogeneous volume (sec.~\ref{SecMFP}) which can be considered as the flow equation for the high temperature homogeneous deformation regime. More generally, it will be shown that the same central results of paper I are also obtained using a log-normal distribution of barrier energies, instead of a normal distribution that was originally employed.

\section{The plastic transition rate via an analogy to statistical mechanics} \label{SecTheory}

The central mathematical challenge is to evaluate the final factor of eqn.~\ref{EqnPlasticRate}:
\begin{equation}
M\left\langle\exp\left(-\frac{E}{k_{\mathrm{B}}T}\right)\right\rangle, \label{EqnPlasticRate1}
\end{equation}
or, with respect to the barrier energy distribution,
\begin{equation}
M\int_{0}^{\infty}dE\,P(E)\exp\left(-\frac{E}{k_{\mathrm{B}}T}\right). \label{EqnGF}
\end{equation}
The integral in eqn.~\ref{EqnGF} is the generating function of the distribution, and for the Gaussian distribution considered in paper I, has a simple closed-form representation. For the log-normal distribution, such a straight forward avenue of calculation appears not to be available, however, the analogy to statistical mechanics derived in paper I provides a procedure for the evaluation of eqn.~\ref{EqnGF}. Indeed, eqn.~\ref{EqnPlasticRate1} can be viewed as
\begin{equation}
M\left\langle\exp\left(-\frac{E}{k_{\mathrm{B}}T}\right)\right\rangle=
\left\langle\sum_{i=1}^{M}\exp\left(-\frac{E_{i}}{k_{\mathrm{B}}T}\right)\right\rangle,\label{EqnPFApp}
\end{equation}
where $\langle\cdots\rangle$ is an average over heterogeneous volumes. The right-hand-side term has the structure of an environmentally averaged partition function average allowing the mathematical problem to exploit the full apparatus of equilibrium statistical mechanics. To obtain a closed form expression of such a partition function, the micro-canonical approach of Derrida~\cite{Derrida1980} can be used to construct the corresponding free (barrier) energy. This procedure is now applied when the barrier energy distribution is taken as a log-normal distribution.

The average number of barrier energies between $E$ and $E+dE$, $\langle\Omega(E)\rangle$, is given by 
\begin{equation}
\langle\Omega(E)\rangle=M(E)dE=MP(E)dE, \label{EqnNumDen}
\end{equation}
where $dE$ must be small enough to ensure a well defined barrier energy but also large enough so that $\langle\Omega(E)\rangle$ is a smooth function of $E$ (see ref. \cite{Derrida1980} for a related discussion). For a sufficiently large heterogeneous volume, fluctuations in eqn.~\ref{EqnNumDen} become small and $\langle\Omega(E)\rangle$ becomes a statistically meaningful quantity, allowing for a corresponding barrier entropy to be defined via $S(E)=k_{\mathrm{B}}\ln\langle\Omega(E)\rangle$. Doing this for the log-normal distribution, eqn.~\ref{EqLND}, results in the barrier entropy:
\begin{equation}
S(E)=k_{\mathrm{B}}\left( \alpha-\frac{\log^{2}\left[\frac{Ea}{\left\langle E\right\rangle}\right]}{4\log a}+\log\left[\frac{1}{\sqrt{2\log a}}\right]+\log\left[\frac{dE}{\sqrt{2\pi}E}\right]\right) \label{EqnSE}.
\end{equation}

To obtain the internal barrier energy and entropy as functions of temperature, an analogy to the thermodynamic definition, 
\begin{equation}
\frac{dS(E)}{dE}=\frac{1}{T}, \label{EqnTER}
\end{equation}
is applied to eqn.~\ref{EqnSE}, giving
\begin{equation}
\log\left[\frac{E(T)a}{\left\langle E\right\rangle}\right]=-2\log a\left(1+\frac{E(T)}{k_{\mathrm{B}}T}\right).
\end{equation}
A solution to the above with respect to $E(T)$ is obtained via the Lambert W-function giving
\begin{equation}
E(T)=k_{\mathrm{B}}T\frac{1}{2\log a}W\left[\frac{\left\langle E\right\rangle}{k_{\mathrm{B}}T}\frac{2\log a}{a^3}\right]. \label{EqnET}
\end{equation}
Eqn.~\ref{EqnET} gives the internal barrier energy as a function of temperature and its substitution into eqn.~\ref{EqnSE} gives the barrier entropy as a function of temperature:
\begin{equation}
S(T)=k_{\mathrm{B}}\left(\alpha-\frac{1}{2}\left(\log\left[2\sqrt{a}\log a\right]+
3\log\left[\frac{E(T)}{\left\langle E\right\rangle}\right]+
\frac{1}{2\log a}\log\left[\frac{E(T)}{\left\langle E(T)\right\rangle}\right]^{2}\right)\right).\label{EqnST}
\end{equation}
In the above, the term containing $dE$ in eqn.~\ref{EqnSE} has been taken as $\ln[dE/(\sqrt{2\pi}E)]=\ln[\left\langle E\right\rangle/E]+\ln[dE/(\sqrt{2\pi}\left\langle E\right\rangle)]\sim\ln[\left\langle E\right\rangle/(\sqrt{2\pi}E)]$. This approximation will be discussed in the proceeding paragraphs.

Eqns.~\ref{EqnET} and \ref{EqnST} give the free barrier entropy, $F(T)=E(T)-TS(T)$, and therefore the final form of eqn.~\ref{EqnPFApp} as
\begin{equation}
M\left\langle\exp\left(-\frac{E}{k_{\mathrm{B}}T}\right)\right\rangle=\exp\left[-\frac{F(T)}{k_{\mathrm{B}}T}\right].\label{EqnPFApp1}
\end{equation}
Subsitution of this back into eqn.~\ref{EqnPlasticRate} gives
\begin{equation}
\left[\tau_{\mathrm{p}}\right]^{-1}=\left[\tau_{\mathrm{p00}}\right]^{-1}\exp\left(\frac{S(T)}{k_{\mathrm{B}}}\right)\exp\left(-\frac{E_{\mathrm{p00}}+E(T)}{k_{\mathrm{B}}T}\right) \label{EqnPlasticRateApp}
\end{equation}
where $E_{\mathrm{p00}}+E(T)$ can be viewed as the apparent barrier energy and the exponential involving $S(T)$ as the number of thermally accessible structural transformations available to the heterogeneous volume. Using eqns.~\ref{EqnE0} to \ref{Eqna}, the $N\rightarrow\infty$ limit of $F(T)$, gives
\begin{equation}
\lim_{N\rightarrow\infty}\frac{F(T)}{N}=k_{\mathrm{B}}T\left(\frac{{E}_{0}^{2}}{\delta\overline{E}_{0}^{2}}\frac{1}{2}\left(\log\left[\frac{k_{\mathrm{B}}T\overline{E}_{0}}{\delta\overline{E}_{0}^{2}}W\left[\frac{\delta\overline{E}_{0}^{2}}{k_{\mathrm{B}}T\overline{E}_{0}}\right]\right]^{2}+2W\left[\frac{\delta\overline{E}_{0}^{2}}{k_{\mathrm{B}}T\overline{E}_{0}}\right]\right)-\overline{\alpha}\right)\label{EqnFBEN}
\end{equation}
demonstrating that for large enough heterogeneous volumes the free barrier energy is an extensive quantity as is the case of the a Gaussian distribution. Indeed by expanding those terms involving the Lambert W-function as a Taylor series to second order in $\delta\overline{E}_{0}^{2}/(k_{\mathrm{B}}T\overline{E}_{0})$ recovers a central result of paper I:
\begin{equation}
\lim_{N\rightarrow\infty}\frac{F(T)}{N}\simeq k_{\mathrm{B}}T\left(\frac{\overline{E}_{0}}{k_{\mathrm{B}}T}-\left(\frac{\delta\overline{E}_{0}}{k_{\mathrm{B}}T}\right)^{2}\right)-k_{\mathrm{B}}T\overline{\alpha}.  \label{EqnFBENLimit}
\end{equation}
That is, for a narrow enough distribution (set by $\delta\overline{E}_{0}$) the free barrier energy limits to that obtained for a Gaussian distribution of barrier energies as derived in Paper I. This is compatible with the fact that for a small enough $\delta\overline{E}_{0}$, the log normal distribution limits to a Gaussian distribution. Eqn.~\ref{EqnFBENLimit} has the virtue that it leads to a simple analytic solution to $F(T_{\mathrm{c}})=0$ for the critical temperature at which significant plastic activity can be expected (see eqn.~11 of ref.~\cite{Derlet2013}).

Eqn.~\ref{EqnFBEN} will be used for further development since it is only at this limit that the right most term in eqn.~\ref{EqnSE} becomes negligible and can be ignored. Indeed Derrida~\cite{Derrida1980} has argued that $dE\sim N^{\zeta}$ where $\zeta<1$. Moreover eqn.~\ref{EqnFBEN} affords a simpler mathematical formalism which is more easily comparable to the normal distribution used in paper I. Reabsorbing the factor $N$ into the free barrier energy gives $F(T)=E(T)-TS(T)$ as
\begin{equation}
F(T)=k_{\mathrm{B}}T\left(\frac{E_{0}^2}{\delta E_{0}^{2}}\frac{1}{2}\left(\log\left[\frac{k_{\mathrm{B}}TE_{0}}{\delta E_{0}^{2}}W\left[\frac{\delta E_{0}^{2}}{k_{\mathrm{B}}T E_{0}}\right]\right]^{2}+2W\left[\frac{\delta E_{0}^{2}}{k_{\mathrm{B}}TE_{0}}\right]\right)-\alpha\right)\label{EqnFBEN1}
\end{equation}
with
\begin{equation}
E(T)=\frac{k_{\mathrm{B}}TE_{0}^{2}}{\delta E_{0}^{2}}W\left[\frac{\delta E_{0}^{2}}{k_{\mathrm{B}}T E_{0}}\right] \label{EqnBEN1}
\end{equation}
and
\begin{equation}
S(T)=k_{\mathrm{B}}\left(\alpha-\frac{E_{0}^{2}}{\delta E_{0}^{2}}\frac{1}{2}\log\left[\frac{k_{\mathrm{B}}TE_{0}}{\delta E_{0}^{2}}W\left[\frac{\delta E_{0}^{2}}{k_{\mathrm{B}}T E_{0}}\right]\right]^{2}\right). \label{EqnBSN1}
\end{equation}

Thus the plastic strain rate (eqn.~\ref{EqnPlasticRate}) becomes
\begin{equation}
\left[\tau_{\mathrm{p}}\right]^{-1}=\left[\tau_{\mathrm{p00}}\right]^{-1}\exp\left(-\frac{E_{\mathrm{p00}}+F(T)}{k_{\mathrm{B}}T}\right) \label{EqnPlasticRate2}
\end{equation}
withe $F(T)$ being given by eqn.~\ref{EqnFBEN1}. The above expression gains its external shear stress dependence via the assumed quadratic dependence of the variance of the barrier energy distribution~\cite{Derlet2013}:
\begin{equation}
\delta E_{0}(\sigma)=\delta E_{0}(\sigma=0)\left(1+\left(\frac{\sigma}{\sigma_{0}}\right)^{2}\right),
\end{equation}
where $\sigma_{0}$ becomes a fitting parameter. Presently, the alternative form
\begin{equation}
\delta E_{0}(\sigma)=\delta E_{0}(\sigma=0)+\delta E_{\sigma}\left(\frac{\sigma}{G}\right)^{2}, \label{EqSD}
\end{equation}
will be used. Here $G$ is an appropriate value for the bulk shear modulus and $\delta E_{\sigma}$ is the corresponding fitting parameter with units of barrier energy.

\section{Macroscopic plasticity --- a mean field approach} \label{SecMFP}

Paper I considered only the characteristic escape rate from a mega-basin. For plastic strain evolution the new mega-basin into which the material enters is of importance. Such a structural transformation from one mega basin to another will contribute to the bulk plastic strain. A natural consequence of a distribution of barrier energies is that any such event will have an associated plastic strain signature which, like the barrier energy, is a random variable. Under zero load conditions, the material traverses (at a long enough time-scale and high enough temperature) a series of mega-basins and the internal plastic strain fluctuates accordingly. Assuming the heterogeneous volumes interact via elasticity, this strain fluctuation corresponds to an internal stress fluctuation and therefore an elastic potential energy landscape.

It is such a coarse grained potential energy landscape which underlies the alpha energy landscape and therefore the heterogeneity of the material. Due to the assumed exponential number of available structural transformations there will also be an exponential number of mega-basins that the material can enter. Hence upon exciting one mega-basin and entering another, the statistics will be negligibly affected, with the consequence being that the system has again access to all possible strain states. Which one it chooses will be stochastically biased towards those that minimise the local elastic energy. This situation does not change upon application of a load, however those structural transformations that can reduce the elastic energy of a particular deformation geometry will become more likely.

So far, the average properties of the non-interacting heterogeneous volumes have been considered, giving $\tau^{-1}_{\mathrm{p}}(T,\sigma^{\mu\nu})$ where $\sigma^{\mu\nu}$ is the externally applied pure shear stress tensor. From the above discussion, the primary interaction between heterogeneous volumes will be of elastic origin. The inclusion of such interactions may be done via the Eshelby construction. Here the goal is to calculate the elastic energy contribution arising from a local plastic strain increment within a particular heterogeneous volume (the $n$th) of volume $V_{0}$. In the Eshelby construction~\cite{Eshelby1957}, this is done by removing from the the amorphous matrix the heterogeneous volume in which the plastic event has occurred, and allowing it to relax to a stress free state given by $\varepsilon_{\mathrm{T}}^{\mu\nu}$. The heterogeneous volume is then deformed elastically to return it to its original shape via the corresponding elastic strain $-\varepsilon_{\mathrm{T}}^{\mu\nu}$ and re-inserted into the matrix. At this point, the particular heterogeneous volume (the inclusion) has a stress derived from Hookes law equal to $\sigma_{\mathrm{T}}^{\mu\nu}$ whilst the surrounding matrix is stress free. The combined system is then allowed to relax to an equilibrium internal stress configuration reducing the inclusion stress by $\sigma_{\mathrm{C}}^{\mu\nu}$. Eshelby has then shown that the total change in elastic energy is given by $V_{0}/2\varepsilon_{\mathrm{T}}^{\mu\nu}(\sigma_{\mathrm{T}}^{\mu\nu}-\sigma_{\mathrm{C}}^{\mu\nu}-\sigma_{\mathrm{E},n}^{\mu\nu})$ where $\sigma_{\mathrm{E},n}^{\mu\nu}$ is an additional external stress (to the $n$th heterogeneous volume). Here, and throughout the text, repeated indices are summed. 

Assuming that this change in elastic energy modifies linearly the intrinsic barrier energy corresponding to this plastic event, one has
\begin{equation}
E_{ni}\rightarrow E_{ni}+\frac{1}{2}V_{0}\varepsilon_{\mathrm{T}}^{\mu\nu}\left(\sigma_{\mathrm{T}}^{\mu\nu}-\sigma_{\mathrm{C}}^{\mu\nu}-\sigma_{\mathrm{E},n}^{\mu\nu}\right).
\end{equation}
The strain rate for the $n$th heterogeneous volume then becomes
\begin{equation}
\dot{\varepsilon}^{\mu\nu}_{n}\left(T,\sigma_{\mathrm{E},n}^{\mu\nu}\right)=\sum_{i=1}^{M}\varepsilon_{\mathrm{T},ni}^{\mu\nu}\left[\tau_{\mathrm{p0},ni}\right]^{-1}\exp\left[-\frac{E_{n,i}+\frac{1}{2}V_{0}\varepsilon_{\mathrm{T},ni}^{\mu\nu}\left(\sigma_{\mathrm{T},ni}^{\mu\nu}-\sigma_{\mathrm{C},ni}^{\mu\nu}-\sigma_{\mathrm{E},n}^{\mu\nu}\right)}{k_{\mathrm{B}}T}\right], \label{EqnPSR}
\end{equation}
where $\varepsilon^{\mu\nu}_{\mathrm{T},ni}$ is the strain increment associated with the barrier energy $E_{n,i}$ and attempt rate $\nu_{n,i}$. Thus each barrier energy is modified from its (local) zero stress value by the total elastic strain energy associated with the plastic event occurring. The strain rate for the RVE is therefore
\begin{equation}
\dot{\varepsilon}^{\mu\nu}_{\mathrm{RVE}}\left(T\right)=\frac{V_{0}}{V_{\mathrm{RVE}}}\sum_{n=1}^{N'}\dot{\varepsilon}^{\mu\nu}_{n}\left(T,\sigma^{\mu\nu}_{n}\right). \label{EqnAvePSR}
\end{equation}
Here $V_{\mathrm{RVE}}$ is the volume of the RVE volume giving $V_{0}/V_{\mathrm{RVE}}=1/N'$ and
\begin{equation}
\dot{\varepsilon}^{\mu\nu}_{\mathrm{RVE}}\left(T\right)=\frac{1}{N'}\sum_{n=1}^{N'}\sum_{i=1}^{M}\varepsilon_{\mathrm{T},ni}^{\mu\nu}\left[\tau_{\mathrm{p0},ni}\right]^{-1}\exp\left[-\frac{E_{n,i}+\frac{1}{2}V_{0}\varepsilon_{\mathrm{T},ni}^{\mu\nu}\left(\sigma_{\mathrm{T},ni}^{\mu\nu}-\sigma_{\mathrm{C},ni}^{\mu\nu}-\sigma_{\mathrm{E},n}^{\mu\nu}\right)}{k_{\mathrm{B}}T}\right]. \label{EqnAvePSR0}
\end{equation}

The elastic relaxation associated with the $i$th plastic event, within and surrounding the $n$th heterogeneous volume, results in $\sigma_{\mathrm{C},ni}^{\mu\nu}$ and an extended stress field field felt by other nearby and distant heterogeneous volumes (through their own $\sigma_{\mathrm{E}}^{\mu\nu}$). From this latter perspective $\sigma_{\mathrm{C},ni}^{\mu\nu}$ can be viewed as a self stress contribution to the elastic energy of the plastic event. The spatial dependence of these stress contributions depend in a non-trivial way on $\sigma_{\mathrm{T}}^{\mu\nu}$ and must be calculated numerically (see for example the book of T. Mura~\cite{MuraBook} and refs.~\cite{Eshelby1957,Bulatov1994a}). It is in this way, that elastic interactions between heterogeneous volumes may be formally introduced. In addition, the $\sigma_{\mathrm{E},n}^{\mu\nu}$ may contain a global applied external stress component written as $\sigma^{\mu\nu}$ giving $\dot{\varepsilon}^{\mu\nu}_{\mathrm{RVE}}\left(T\right)=\dot{\varepsilon}^{\mu\nu}_{\mathrm{RVE}}\left(T,\sigma^{\mu\nu}\right)$.

Thus for each heterogeneous volume, $\sigma_{\mathrm{E},n}^{\mu\nu}$ arises from an external applied stress plus an internal stress arising from the elastic interaction between the plastic events from all other heterogeneous volumes. Such a formalism has been considered by Bulatov and Argon~\cite{Bulatov1994a,Bulatov1994b,Bulatov1994c} in their development of a thermal activation model for the plasticity of glasses, and more recently using the finite-element-method by Homer and Schuh~\cite{Homer2009,Homer2010,Li2013}. In the present work a mean field approach will be taken which exploits the fact that the local contribution to $\sigma_{\mathrm{E},n}^{\mu\nu}$ arising from other heterogeneous volumes will on average equal zero, giving $\sigma_{\mathrm{E},n}^{\mu\nu}=\sigma^{\mu\nu}$. For an assumed isotropic elasticity, $\varepsilon_{\mathrm{T},ni}^{\mu\nu}(\sigma_{\mathrm{T},ni}^{\mu\nu}-\sigma_{\mathrm{C},ni}^{\mu\nu})$ will equal $4V_{0}(G-\delta G_{ni})\gamma_{\mathrm{T},ni}^{2}$ where $\gamma_{\mathrm{T},ni}$ will be the shear strain magnitude of the plastic event and $G$ is the appropriate bulk isotropic shear modulus for the system. $\delta G_{ni}$ embodies the elastic relaxation of the plastic strain event and will be different for each plastic event. Within the current mean field approach this will be absorbed into $G$ resulting in the $G$ being an effective bulk shear modulus of the system.

These approximations afford a simple average with respect to heterogeneous volumes resulting in eqn.~\ref{EqnAvePSR0} becoming
\begin{equation}
\dot{\varepsilon}^{\mu\nu}_{\mathrm{RVE}}\left(T,\sigma^{\mu\nu}\right)=M\left\langle\varepsilon_{\mathrm{T}}^{\mu\nu}\left[\tau_{\mathrm{p0}}\right]^{-1}\exp\left[-\frac{E_{\mathrm{p0}}+2V_{0}G\gamma_{\mathrm{T}}^{2}-\frac{1}{2}V_{0}\varepsilon_{\mathrm{T}}^{\mu\nu}\sigma^{\mu\nu}}{k_{\mathrm{B}}T}\right]\right\rangle,
\end{equation}
in which the average is performed with respect to $\varepsilon_{\mathrm{T}}^{\mu\nu}$, $\tau_{\mathrm{p0}}$ and $E_{\mathrm{p0}}$. Since there should be no correlation between the strain increment $\varepsilon_{\mathrm{T}}^{\mu\nu}$ and the parameters $\tau_{\mathrm{p0}}$ and $E_{\mathrm{p0}}$ associated with the escape of a particular mega-basin (since each strain state is always accessible), the above average with respect to $\varepsilon_{\mathrm{T}}^{\mu\nu}$ may be performed independently from $\tau_{\mathrm{p0}}$ and $E_{\mathrm{p0}}$, giving,
\begin{eqnarray}
\dot{\varepsilon}^{\mu\nu}_{\mathrm{RVE}}\left(T,\sigma^{\mu\nu}\right)&=&\left\langle\varepsilon_{\mathrm{T}}^{\mu\nu}\exp\left[\frac{-2V_{0}G\gamma_{\mathrm{T}}^{2}+\frac{1}{2}V_{0}\varepsilon_{\mathrm{T}}^{\mu\nu}\sigma^{\mu\nu}}{k_{\mathrm{B}}T}\right]\right\rangle_{\varepsilon_{\mathrm{T}}^{\mu\nu}} M\left\langle\left[\tau_{\mathrm{p0}}\right]^{-1}\exp\left[-\frac{E_{\mathrm{p0}}}{k_{\mathrm{B}}T}\right]\right\rangle_{\tau_{\mathrm{p0}},E_{\mathrm{p0}}} \nonumber \\
&=&\left\langle\varepsilon_{\mathrm{T}}^{\mu\nu}\exp\left[\frac{-2V_{0}G\gamma_{\mathrm{T}}^{2}+\frac{1}{2}V_{0}\varepsilon_{\mathrm{T}}^{\mu\nu}\sigma^{\mu\nu}}{k_{\mathrm{B}}T}\right]\right\rangle_{\varepsilon_{\mathrm{T}}^{\mu\nu}}\left[\tau_{\mathrm{p}}\right]^{-1}\left(T,\sigma^{\mu\nu}\right).\label{EqnAvePSR1}
\end{eqnarray}
In the last equality, $\left[\tau_{\mathrm{p}}\right]^{-1}\left(T,\sigma^{\mu\nu}\right)$ is given by eqn.~\ref{EqnPlasticRate}. 

A consequence of such a lack of correlation is that the magnitude of a barrier energy is, on average, not correlated with the magnitude of the resulting plastic strain increment. This scenario, again, appeals to the notion of an exponentially large diversity of possible structural excitation. Moreover, numerical evidence in simple Lennard Jones structural glasses exists, demonstrating that the strain associated in traversing a nearby local PEL minima does not at all correlate with the height of the intermediate saddle-point (barrier) energy~\cite{Rodney2009a,Rodney2009b}.

The plastic strain average in eqn.~\ref{EqnAvePSR1} may be written as
\begin{equation}
\left\langle\varepsilon_{\mathrm{T}}^{\mu\nu}\exp\left[\frac{-2V_{0}G\gamma_{\mathrm{T}}^{2}+\frac{1}{2}V_{0}\varepsilon_{\mathrm{T}}^{\mu\nu}\sigma^{\mu\nu}}{k_{\mathrm{B}}T}\right]\right\rangle_{\varepsilon_{\mathrm{T}}^{\mu\nu}}=\frac{\partial}{\partial\left[\frac{\frac{1}{2}V_{0}\sigma^{\mu\nu}}{k_{\mathrm{B}}T}\right]}\left\langle\exp\left[\frac{-2V_{0}G\gamma_{\mathrm{T}}^{2}+\frac{1}{2}V_{0}\varepsilon_{\mathrm{T}}^{\mu\nu}\sigma^{\mu\nu}}{k_{\mathrm{B}}T}\right]\right\rangle_{\varepsilon_{\mathrm{T}}^{\mu\nu}}\label{EqnGFIdentity0}
\end{equation}
and therefore the goal is to calculate
\begin{equation}
\left\langle\exp\left[\frac{-2V_{0}G\gamma_{\mathrm{T}}^{2}+\frac{1}{2}V_{0}\varepsilon_{\mathrm{T}}^{\mu\nu}\sigma^{\mu\nu}}{k_{\mathrm{B}}T}\right]\right\rangle_{\varepsilon_{\mathrm{T}}^{\mu\nu}}. \label{EqnStrainAv}
\end{equation}
To evaluate this average further, the two dimensional case of plane strain will be considered involving only an external pure shear stress component:
\begin{equation}
\sigma^{\mu\nu}=\left(
  \begin{array}{ccc}
    0 & \sigma & 0 \\
    \sigma & 0 & 0 \\
    0 & 0 & 0 \\
  \end{array}
\right).
\end{equation}
Under these constraints, the characteristic shear strain matrix takes the form
\begin{equation}
\gamma_{\mathrm{T}}\left(
  \begin{array}{ccc}
    2\sin\theta_{\mathrm{T}}\cos\theta_{\mathrm{T}} & \cos2\theta_{\mathrm{T}} & 0 \\
    \cos2\theta_{\mathrm{T}} & -2\sin\theta_{\mathrm{T}}\cos\theta_{\mathrm{T}} & 0 \\
    0 & 0 & 0 \\
  \end{array}
\right)
\end{equation}
giving
\begin{equation}
\varepsilon_{\mathrm{T}}^{\mu\nu}\sigma^{\mu\nu}=2\gamma_{\mathrm{T}}\sigma\cos2\theta_{\mathrm{T}}.
\end{equation}
Thus all possible plastic strain increments are defined by an orientation angle $\theta_{\mathrm{T}}$ and a plastic shear strain magnitude $\gamma_{\mathrm{T}}$. Due to the exponentially large number of plastic strain increments available and the lack of a preferred direction, the random variable $\theta_{\mathrm{T}}$ is assumed to be derived from a uniform distribution ranging from $\theta_{\mathrm{T}}=0$ to $\theta_{\mathrm{T}}=\pi$. On the other hand, for reasonable values of the shear strain magnitude, the distribution for $\gamma_{\mathrm{T}}$ is expected to be bounded. Presently (and for mathematical simplicity) the distribution is assumed to be a normal distribution centred on zero with standard deviation $\delta\gamma$. It is acknowledged that other distributions are also feasible. With these assumptions eqn.~\ref{EqnStrainAv} becomes
\begin{equation}
\left\langle\exp\left[\frac{-2V_{0}G\gamma_{\mathrm{T}}^{2}+\frac{1}{2}V_{0}\varepsilon_{\mathrm{T}}^{\mu\nu}\sigma^{\mu\nu}}{k_{\mathrm{B}}T}\right]\right\rangle_{\theta_{\mathrm{T}},\gamma_{\mathrm{T}}}=\exp\left[\left(\frac{V_{0}\delta\gamma'\sigma}{2k_{\mathrm{B}}T}\right)^{2}\right]\mathrm{I}_{0}\left[\left(\frac{V_{0}\delta\gamma'\sigma}{2k_{\mathrm{B}}T}\right)^{2}\right]. \label{EqnGFIdentity1}
\end{equation}
where $\mathrm{I}_{n}(x)$ is the modified Bessel function of order $n$ and
\begin{equation}
\left(\delta\gamma'\right)^{2}=\frac{\delta\gamma^{2}}{1+\frac{4V_{0}G\delta\gamma^{2}}{k_{\mathrm{B}}T}}\simeq\delta\gamma^{2}. \label{EqnApprox}
\end{equation}
In sec.~\ref{SecExp}, the latter approximation will be shown to be valid for temperatures close to the glass transition temperature.

Using eqn.~\ref{EqnGFIdentity0} this finally gives the strain rate per heterogeneous volume as
\begin{eqnarray}
\dot{\gamma}_{\mathrm{RVE}}\left(T,\sigma\right)&=&2\left(\frac{V_{0}\delta\gamma^{2}\sigma}{2k_{\mathrm{B}}T}\right)\exp\left[\left(\frac{V_{0}\delta\gamma\sigma}{2k_{\mathrm{B}}T}\right)^{2}\right]\nonumber \\
& &\times\left(\mathrm{I}_{0}\left[\left(\frac{V_{0}\delta\gamma\sigma}{2k_{\mathrm{B}}T}\right)^{2}\right]+\mathrm{I}_{1}\left[\left(\frac{V_{0}\delta\gamma\sigma}{2k_{\mathrm{B}}T}\right)^{2}\right]\right)\nonumber \\
& &\times\left[\tau_{\mathrm{p}}\right]^{-1}\left(T,\sigma^{\mu\nu}\right)
\end{eqnarray}
Using eqn.~\ref{EqnPlasticRate2} and writing, $\delta\gamma=\delta\gamma_{0}/V_{0}$, this reduces to
\begin{eqnarray}
\dot{\gamma}_{\mathrm{RVE}}\left(T,\sigma\right)&=&\delta\dot{\gamma_{0}}\left(\frac{\delta\Omega\sigma}{k_{\mathrm{B}}T}\right)\exp\left[\left(\frac{\delta\Omega\sigma}{2k_{\mathrm{B}}T}\right)^{2}\right]\left(\mathrm{I}_{0}\left[\left(\frac{\delta\Omega\sigma}{2k_{\mathrm{B}}T}\right)^{2}\right]+\mathrm{I}_{1}\left[\left(\frac{\delta\Omega\sigma}{2k_{\mathrm{B}}T}\right)^{2}\right]\right)\nonumber \\
& &\times\exp\left[-\frac{E_{\mathrm{p00}}+F(T,\sigma)}{k_{\mathrm{B}}T}\right] \label{EqnAvePSR3}
\end{eqnarray}
where
\begin{equation}
\delta\Omega=\frac{\delta\gamma_{0}}{2} \label{EqnActivationVolume}
\end{equation}
and
\begin{equation}
\delta\dot{\gamma}_{0}=\frac{2\delta\gamma_{0}}{V_{0}\tau_{\mathrm{p00}}}.\label{EqnCharacteristicSR}
\end{equation}
Eqn.~\ref{EqnAvePSR3}, written explicitly in terms of the internal barrier energy and barrier entropy, becomes
\begin{eqnarray}
\dot{\gamma}_{\mathrm{RVE}}\left(T,\sigma\right)&=&\delta\dot{\gamma_{0}}\exp\left(\frac{S(T,\sigma)}{k_{\mathrm{B}}T}\right)\nonumber\\
& &\times\left(\frac{\delta\Omega\sigma}{k_{\mathrm{B}}T}\right)\exp\left[\left(\frac{\delta\Omega\sigma}{2k_{\mathrm{B}}T}\right)^{2}\right]\left(\mathrm{I}_{0}\left[\left(\frac{\delta\Omega\sigma}{2k_{\mathrm{B}}T}\right)^{2}\right]+\mathrm{I}_{1}\left[\left(\frac{\delta\Omega\sigma}{2k_{\mathrm{B}}T}\right)^{2}\right]\right) \nonumber \\
& &\times\exp\left[-\frac{E_{\mathrm{p00}}+E(T,\sigma)}{k_{\mathrm{B}}T}\right].\label{EqnAvePSR4}
\end{eqnarray}
 
The above equation should be compared to the early thermal activation models grounded on the work of Spaepen~\cite{Spaepen1977} and Argon~\cite{Argon1979}, which for homogeneous plastic deformation in the vicinity of the glass transition temperature is usually taken as~\cite{Schuh2007,Wang2011}
\begin{equation}
\dot{\gamma}=\alpha_{\mathrm{ta}}\nu_{\mathrm{ta}}\gamma_{\mathrm{ta}}\exp\left[-\frac{Q_{\mathrm{ta}}}{k_{\mathrm{B}}T}\right]\sinh\left[\frac{V_{\mathrm{ta}}\sigma}{k_{\mathrm{B}}T}\right]
 \label{EqnSimpleT}
\end{equation}
where $\alpha_{\mathrm{ta}}$ contains various quantities including the fraction of volume available to deformation, $\nu_{\mathrm{ta}}$ is the attempt frequency, $\gamma_{\mathrm{ta}}$ is the characteristic strain, $Q_{\mathrm{ta}}$ is the activation energy and $V_{\mathrm{ta}}$ is the activation volume. Comparison of eqn.~\ref{EqnAvePSR4} to \ref{EqnSimpleT} suggests that $\delta\dot{\gamma}_{0}$ can be viewed as a characteristic strain rate ($=\nu_{\mathrm{ta}}\gamma_{\mathrm{ta}}$ of eqn.~\ref{EqnSimpleT}), $\exp\left[S(T,\sigma)/k_{\mathrm{B}}\right]$ as a temperature and stress dependent number of available structural transformations, $E_{\mathrm{p00}}+E(T,\sigma)$ as a temperature and stress dependent activation energy, and $\delta\Omega$ as an activation volume. This latter interpretation is motivated by the term $x\exp(x^2)(\mathrm{I}_{0}[x^{2}]+\mathrm{I}_{1}[x^{2}])$ being qualitatively similar to $\sinh(x)$, with both limiting to a linear function for small $x$.

Thus, although the present theory has a number of quite different physical assumptions when compared to earlier high-temperature thermal activation models, the mean-field prediction of the average strain rate as a function of temperature and stress in the high temperature/low strain rate regime does not operationally differ from these early models. This observation will be shown in more detail in the next section, when the model is fitted to actual experimental data.

\section{Application to the high temperature deformation data of Lu, Ravichandran and Johnson~\cite{Johnson2005}} \label{SecExp}

Inspection of eqn.~\ref{EqnAvePSR3} reveals the free parameters of the model are $\delta\dot{\gamma}_{0}$, $\delta\Omega$, and those associated with $E_{\mathrm{p00}}+F(T)$ which are $E_{\mathrm{p00}}$, $E_{0}$, $\delta E_{0}$, $\alpha$ and $\delta E_{\sigma}$. The latter shear stress dependence gets its definition via eqn.~\ref{EqSD} where here $\delta E_{0}$ is equal to $\delta E_{0}(\sigma=0)$. As in paper I, $E_{0}$ and $\delta E_{0}$, which define the barrier energy distribution can be entirely determined from experimentally accessible parameters, in particular via the viscosity $\eta_{\mathrm{g}}$ and fragility $m$ at the glass transition temperature $T_{\mathrm{g}}$.

Linear (Newtonian) viscosity may be formally given as
\begin{equation}
\eta_{\mathrm{RVE}}=\lim_{\sigma\rightarrow0}\frac{\sigma}{\dot{\gamma}_{\mathrm{RVE}}}=\frac{k_{\mathrm{B}}T}{\delta\Omega\delta\dot{\gamma}_{0}}\exp\left[\frac{E_{\mathrm{p00}}+F(T,\sigma=0)}{k_{\mathrm{B}}T}\right].\label{EqnViscocity}
\end{equation}
One common method to define the glass transition temperature is when the viscosity reaches $\eta_{\mathrm{g}}=10^{12}$ Pa-sec at a cooling rate of 20 K/min. That is,
\begin{equation}
\eta_{\mathrm{g}}=\frac{k_{\mathrm{B}}T_{\mathrm{g}}}{\delta\Omega\delta\dot{\gamma}_{0}}\exp\left[\frac{E_{\mathrm{p00}}+F(T_{\mathrm{g}},\sigma=0)}{k_{\mathrm{B}}T_{\mathrm{g}}}\right].\label{EqnViscocity1}
\end{equation}
In addition to the value of the viscosity at $T_{\mathrm{g}}$, a common material parameter to structural glasses is the fragility --- the rate at which viscosity changes with respect to temperature at $T_{\mathrm{g}}$,
\begin{equation}
m=\left.\frac{d\log\eta/\eta_{0}}{d\left(T_{\mathrm{g}}/T\right)}\right|_{T=T_{\mathrm{g}}}. \label{EqnFragility}
\end{equation}
Substitution of eqn.~\ref{EqnViscocity} into eqn.~\ref{EqnFragility} gives 
\begin{equation}
m\ln10+1=\frac{E_{\mathrm{p00}}+E(T_{\mathrm{g}},\sigma=0)}{k_{\mathrm{B}}T_{\mathrm{g}}}
\end{equation}
or the internal barrier energy at $T_{\mathrm{g}}$
\begin{equation}
E(T_{\mathrm{g}},\sigma=0)=k_{\mathrm{B}}T_{\mathrm{g}}\left(m\ln10+1\right)-E_{\mathrm{p00}}. \label{EqnConstraint1}
\end{equation}
This physically appealing result is general and arises because eqn.~\ref{EqnTER} requires $T\partial S/\partial T=\partial E/\partial T$ or equivalently the expression, $U/T^{2}=\partial(F/T)/\partial T$, which is analogous to the Gibbs-Helmholtz equation in thermodynamics. Further substitution of eqn.~\ref{EqnConstraint1} into \ref{EqnViscocity1}, gives the barrier entropy at $T_{\mathrm{g}}$
\begin{equation}
S(T_{\mathrm{g}},\sigma=0)=k_{\mathrm{B}}\left(m\ln10+1-\ln\left[\frac{\eta_{\mathrm{g}}\delta\Omega\delta\dot{\gamma}_{0}}{k_{\mathrm{B}}T_{\mathrm{g}}}\right]\right). \label{EqnConstraint2}
\end{equation}
Thus the experimental parameters $\eta_{\mathrm{g}}$, $m$ define the free barrier energy at $T_{\mathrm{g}}$ for a given choice of $\delta\Omega$, $\delta\dot{\gamma}_{0}$ and $\alpha$.
Eqns.~\ref{EqnConstraint1} and \ref{EqnConstraint2} in conjunction with eqns.~\ref{EqnBEN1} and \ref{EqnBSN1} also uniquely determine $E_{0}$ and $\delta E_{0}$,
\begin{equation} 
E_{0}=E(T_{\mathrm{g}},\sigma=0)\exp\left[-\frac{2\left(S(T_{\mathrm{g}},\sigma=0)-k_{\mathrm{B}}\alpha\right)}{E(T_{\mathrm{g}},\sigma=0)}\right]
\end{equation}
and
\begin{equation} 
\delta E_{0}=E_{0}\sqrt{\frac{k_{\mathrm{B}}T_{\mathrm{g}}}{E(T_{\mathrm{g}},\sigma=0)}\log\left[\frac{E_{0}}{E(T_{\mathrm{g}},\sigma=0)}\right]}.
\end{equation}

It is noted that the time rate of change of viscosity at the glass transition is given as $\dot{\eta}_{\mathrm{g}}=\eta_{\mathrm{g}}m\mathrm{CR}/T_{\mathrm{g}}$ where CR is the cooling rate from the under-cooled liquid regime. Thus the cooling rate does not give additional information about the theory and emphasizes the fact that the current thermal activation approach is not applicable to the physics of the under-cooled liquid regime.

The remaining free parameters are $\delta\dot{\gamma}_{0}$, $\delta\Omega$, $E_{\mathrm{p00}}$, $\alpha$ and $\delta E_{\sigma}$. $E_{\mathrm{p00}}$ can be determined from dynamical mechanical analysis or differential scanning calorimetry experiments~\cite{Wang2011} and the rest from steady state strain rate versus stress data as a function of temperature, a common data set used to fit to the Arrhenius form of eqn.~\ref{EqnSimpleT}. For Vitreloy-1 such a data set is found in the work of Lu {\em et al}~\cite{Lu2003}. For this structural glass initial estimates of $T_{\mathrm{g}}=623$ K, $m=40$ and $E_{\mathrm{p00}}=1.4$ eV are taken from both Lu {\em et al}~\cite{Lu2003} and Wang~\cite{Wang2011}.

\begin{figure}
\begin{center}
\includegraphics[clip,width=0.9\textwidth]{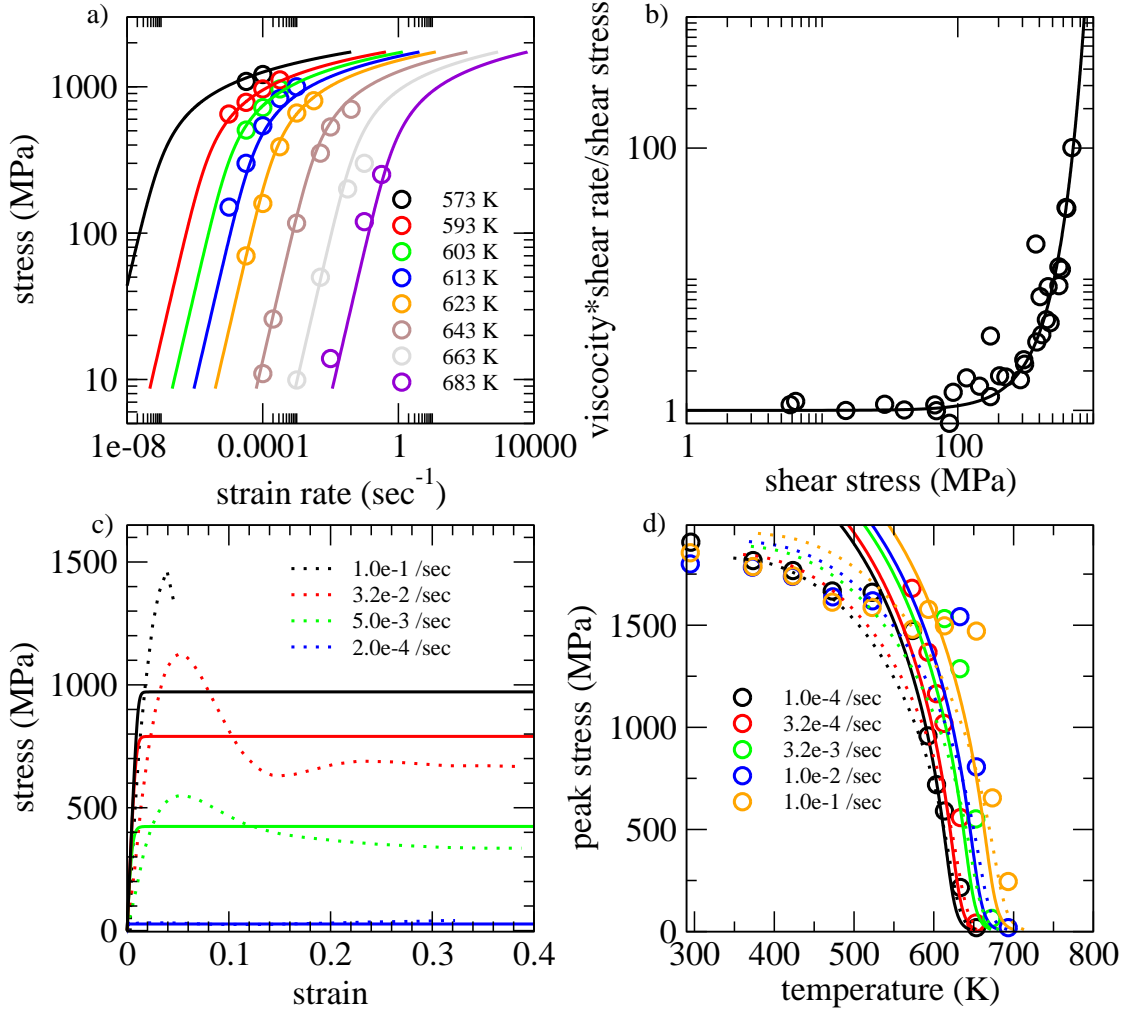}
\end{center}
\caption{For Viteroly-1, data of a) steady state strain rate versus stress, b) Newtonian viscosity multiplied by shear strain rate and divided by shear stress, c) uniaxial stress-strain curves for various strain rates and d) peak stress as a function of temperature. The symbol (or dashed lines in c) data display the experimental data of Lu {\em et al} and the solid curves of similar colour display the corresponding model fit. In d) the dashed curves represent the model trends when fitted directly to the experimental peak stress data of c).} \label{FigExpCurves} 
\end{figure}

Fig.~\ref{FigExpCurves}a plots the experimental steady state strain rate versus stress for a number of temperatures close to and above the glass transition temperature. Following Schuh {\em et al}~\cite{Schuh2007} and Wang~\cite{Wang2011} it is this data to which the current model (eqn.~\ref{EqnAvePSR3}) will be fitted with all parameters allowed to vary. The experimental data is taken from uniaxial compression tests and for conversion between the strain rate and uni-axial stress to the appropriate shear quantities, the former is multiplied by $\sqrt{3}$ and the latter is divided by $\sqrt{3}$. This is motivated by fig.~\ref{FigExpCurves}b, which plots Lu's linear viscosity (fig.~8 of ref.~\cite{Lu2003}) data multiplied by the shear strain rate and divided by the shear stress. Such numerical factors are common and represent an effective angle of dominant slip activity in a uniaxial deformation experiment, see for example Kawamura {\em et al} ~\cite{Kawamura1988}. As expected, for a wide range of low shear stresses, the experimental data in fig.~\ref{FigExpCurves}b has converged to approximately unity indicating a stress independent Newtonian viscosity coefficient. The solid curves in figs.~\ref{FigExpCurves}a-b indicate a fit of the model obtained via the simulated annealing global minimisation algorithm~\cite{Corana1987}. The corresponding parameters of this fit are $T_{\mathrm{g}}\simeq621.5$ K, $m\simeq38.7$, $E_{\mathrm{p00}}\simeq1.44$ eV, $\delta\dot{\gamma}_{0}\simeq2.95\times10^{9}$ sec$^{-1}$, $\delta\Omega\simeq1.99\times10^{-29}$ m$^{3}$, $\alpha\simeq107.6$ and $\delta E_{\sigma}\simeq2.98$ eV, with $\eta_{\mathrm{g}}$ fixed at $10^{12}$ Pa-sec.

It is noted that for these parameters, the numerator of eqn.~\ref{EqnApprox} evaluated at $T_{\mathrm{g}}$ equals $\simeq1.03$ for $V_{0}\simeq10$ nm$^{3}$, justifiying the ensuing approximation of eqn.~\ref{EqnApprox}.

When interpreted as an activation volume, the parameter $\delta\Omega$ has a numerical value that is similar to that found in refs.~\cite{Schuh2007,Wang2011} when using the Arrhenius form of eqn.~\ref{EqnSimpleT}. When assuming that the volume of the heterogeneous volumes is one to three orders of magnitude larger than this ``activation volume'', eqn.~\ref{EqnActivationVolume} and \ref{EqnCharacteristicSR} together with the numerical value of $\delta\dot{\gamma}_{0}$ indicate a value for $\tau_{\mathrm{p00}}$ ranging somewhere between $10^{9}$ to $10^{12}$ --- a range of values compatible with it being some multiple of the Debye frequency of the system. The numerical value of $\delta E_{\sigma}$ indicates a weak stress dependence of the distribution of barrier energies, a consequence of which is that the apparent barrier energy has a weak stress dependence and therefore a value ($E(T_{\mathrm{g}})+E_{\mathrm{p00}}\simeq4.82$ eV) differing little from the activation energy obtained when using eqn.~\ref{EqnSimpleT}, which is $Q\simeq4.6$ eV \cite{Schuh2007,Wang2011}. The corresponding log-normal barrier energy parameters have values of $E_{0}=15.02$ eV and $\delta E_{0}=2.3$ eV indicating that in the high temperature regime, the relevant part of the distribution ($\sim k_{\mathrm{B}}T_{\mathrm{g}}$ remains that of low barrier energy tail.

Fig.~\ref{FigExpCurves}c displays experimental uniaxial deformation curves at a temperature of 643K for a number of different strain rates. The data is taken directly from fig.~2 of ref.~\cite{Lu2003}. The higher strain rate curves are characterised by an initial rise in stress that may be loosely associated with the elastic deformation regime, a peak stress indicating the onset of a material instability and an eventual stead state flow regime. Appendix B of paper I used a simple method to obtain a constant strain rate stress strain curve from eqn.~\ref{EqnAvePSR3}, resulting in the stress at time $t+\Delta t$ being given by
\begin{equation}
\sigma(t+\delta t)=\sigma(t)+Y\left[\dot{\varepsilon}_{\mathrm{total}}\delta t-\int_{t}^{t+\delta t}dt\,\dot{\varepsilon}_{\mathrm{p}}(T,\sigma(t))\right].
\end{equation}
For a small enough time interval this may be approximated as
\begin{equation}
\sigma(t+\delta t)=\sigma(t)+Y\dot{\varepsilon}_{\mathrm{total}}\delta t\left[1-\frac{\dot{\varepsilon}_{\mathrm{p}}(T,\sigma(t))}{\dot{\varepsilon}_{\mathrm{total}}}\right]
\end{equation}
and iterated to generate a stress-strain curve.

In the above the plastic strain rate, $\dot{\varepsilon}_{\mathrm{p}}(T,\sigma)$ is given by $\sqrt{3}\dot{\gamma}_{\mathrm{RVE}}\left(T,\sigma\right)$ (eqn.~\ref{EqnAvePSR3}), $Y$ is the appropriate bulk Youngs modulus and $\dot{\varepsilon}_{\mathrm{total}}$ the chosen constant total strain rate. Using this model, fig.~\ref{FigExpCurves}c also displays the resulting model stress-strain curves. Whilst the steady state stress regime is well described by the model, particularly for low values of strain rate, the peak stress regime associated with the emergence of a stress overshoot is not at all present --- a result due to the mean field nature of the current plasticity model, which necessarily can only describe homogeneous plasticity. Fig.~\ref{FigExpCurves}d now shows the experimental peak stress values of Lu {\em et al} \cite{Lu2003} for different strain rates as a function of temperature (shown as open circles in the figure). With $T_{\mathrm{g}}$ at approximately 623K, such data clearly shows the transition from a high temperature/low strain rate homogeneous plasticity to a low temperature/high strain rate regime of deformation. Whilst the model is unable to reproduce the peak stress behaviour it is revealing to plot on this curve the temperature dependence of the steady state stresses for comparable strain rates (solid curves). For the lowest strain rate, good agreement is seen in the high temperature regime where experimentally the peak stress regime is absent. For larger strain rates, the quantitative agreement reduces due to the mean field nature of the current plasticity model --- however qualitatively the trends are similar. For this very same reason, the rapid change in experimental behaviour seen as the temperature reduces is not evident in the model, although at higher stresses (beyond the range of the figure) the predicted steady state stress does limit to a plateau with a weak temperature dependence --- a regime that was investigated in Paper I (see fig.~8 of~\cite{Derlet2013}).

Having said the above, fig.~\ref{FigExpCurves}d also includes (the dashed lines) the consequences of the model when fitted to the peak stress data, as in paper I, rather than the steady state data. The emerging plateau region at low temperatures now becomes evident, although the sharper transition to this temperature regime seen when using the modified Gaussian of paper I is no longer evident when using the more restrictive log-normal distribution of barrier energies. The parameters of this fit are $T_{\mathrm{g}}\simeq625.0$, $m\simeq35.1$, $E_{\mathrm{p00}}\simeq1.44$ eV, $\delta\dot{\gamma}_{0}\simeq7.15\times10^{9}$ sec$^{-1}$, $\delta\Omega\simeq2.40\times10^{-29}$ m$^{3}$,  $\alpha\simeq119.7$ and $\delta E_{\sigma}\simeq6.25$ eV. These values do not differ substantially from those associated with the fit to steady state flow stress data. The corresponding log-normal barrier energy parameters have values of $E_{0}=32.5$ eV and $\delta E_{0}=6.7$ eV which represents a distribution that is closer in position to the the fitted modified Gaussian distribution used in paper I (see fig.~7c of~\cite{Derlet2013}). Whilst such an approach is certainly not justified in the temperature regime where there is a cross-over from homogeneous to heterogeneous plasticity, and a strong difference between the peak stress and flow stress, well below the glass transition temperature there is little experimental distinction between these stresses. Thus one could argue (as was done in paper I) that mean field would be applicable both to the quite low temperature and high temperature deformation modes for the onset of plastic deformation (that is yield), with the intermediate cross-over regime not being well described. The dashed curves in fig.~\ref{FigExpCurves}d tend to support this conclusion. 

\section{Discussion and concluding remarks} \label{SecConc}

The present thermal activation theory provides a entirely kinetic interpretation of the fragility index at the glass transition temperature. This may be seen by writing the barrier entropy, eqn.~\ref{EqnConstraint2}, as
\begin{equation}
S(T_{\mathrm{g}})+k_{\mathrm{B}}\ln\left[\frac{\eta_{\mathrm{g}}\delta\Omega\delta\dot{\gamma}_{0}}{k_{\mathrm{B}}T_{\mathrm{g}}}\right]=k_{\mathrm{B}}\left(m\ln10+1\right).
\end{equation}
or rather as
\begin{equation}
S(T_{\mathrm{g}})+k_{\mathrm{B}}\ln\left[\frac{\tau_{\mathrm{Exp}}}{\tau_{\mathrm{p00}}}\right]=k_{\mathrm{B}}\left(m\ln10+1\right).\label{EqnConstraint3}
\end{equation}
with $\tau_{\mathrm{Exp}}=\eta_{\mathrm{g}}V_{0}\delta\gamma^{2}/\left(2k_{\mathrm{B}}T_{\mathrm{g}}\right)$ being viewed as the characteristic experimental time scale associated with the deformation experiment (see paper I). If $\tau_{\mathrm{Exp}}$ where equal to the fundamental time scale of the $\beta$-relaxation modes, $\tau_{\mathrm{p00}}$, then $S(T_{\mathrm{g}})$ would equal $m\ln10+1$ where $m$ is the correspondingly measured fragility. In this limit, the fragility is therefore a direct measure of the apparent number of structural transformations available to each heterogeneous volume element at $T_{\mathrm{g}}$, that is, $S(T_{\mathrm{g}})/k_{\mathrm{B}}=\alpha_{\mathrm{App}}(T)=m\ln10+1$. An analogous identification has also been proposed between fragility and the configurational entropy of the under-cooled liquid PEL by Sastry~\cite{Sastry2001}. This present result immediately implies that $\alpha$ cannot be less than $m\ln10+1$ since $S(T)/k_{\mathrm{B}}<\alpha$ (eqn.~\ref{EqnST}). Note that, like the barrier entropy, the fragility is an extensive quantity.

For a realistic deformation experiment, $\tau_{\mathrm{Exp}}>>\tau_{\mathrm{p00}}$, resulting in the above equality between $S(T_{\mathrm{g}})$ and $m$ being an over estimation of the barrier entropy. This is because over the period of time, $\tau_{\mathrm{Exp}}$, significant {\em experimentally unresolvable} activity occurs which also contributes to the correspondingly measured fragility. Through the term $k_{\mathrm{B}}\ln\left[\tau_{\mathrm{Exp}}/\tau_{\mathrm{p00}}\right]$ this activity contributes to an effective reduction of the needed barrier entropy. Thus when the glass transition is viewed as an entirely kinetic phenomenon (as is done here), the fragility becomes a direct measure of the apparent number of available $\alpha$-mode structural transformations. Thus eqn.~\ref{EqnConstraint3} (and eqn.~\ref{EqnConstraint2}) is considered more fundamental than the more familiar eqn.~\ref{EqnConstraint1} which (here) arises from the requirement that the temperature $T$ defined via eqn.~\ref{EqnTER} corresponds to the thermodynamic temperature of the system.

Using the fitted parameters of Viteroly-1 obtained in sec.~\ref{SecExp}, the value of the ratio $\tau_{\mathrm{Exp}}/\tau_{\mathrm{p00}}\simeq3\times10^{13}$ suggests a reasonable choice of $\tau_{\mathrm{p00}}$ (in the range of $10^{-9}$ to $10^{-12}$ sec) will give a $\tau_{\mathrm{Exp}}$ that is within the domain of a typical deformation experiment.

The current mean-field description of macroscopic plasticity is only suited to purely homogeneous plasticity, where within one RVE, it is valid to describe the plasticity as arising from the average response of a heterogeneous volume. Whilst this might be valid for a broad range of temperatures in the flow stress regime, figs.~\ref{FigExpCurves}c-d demonstrate that for high enough strain rate or low enough temperature, the mean field approximation is unable to describe the presumably heterogeneous transition from elasticity to macroscopic plasticity seen in experiment. Indeed at quite low temperatures, where the available number of structural transformations per heterogeneous volume approaches unity a new regime of strongly heterogeneous statistics emerges associated with that of the extreme value. The average temperature at which this occurs is referred to as the kinetic freezing temperature and may be found as the solution to the barrier entropy equalling zero, $S(T_{\mathrm{f}})=0$: the case when there exists, on average, one available structural transformation per heterogeneous volume. Using eqn.~\ref{EqnBSN1} and the known properties of the Lambert W-function, this temperature is found to be
\begin{equation}
T_{\mathrm{f}}(\sigma)=\frac{1}{k_{\mathrm{B}}}\frac{\delta E_{0}(\sigma)}{\sqrt{2\alpha}}\exp\left[-\sqrt{2\alpha}\frac{\delta E_{0}(\sigma)}{E_{0}}\right]. \label{EqnFT}
\end{equation}
for the log-normal distribution. A Gaussian distribution gives a similar result, but without the exponential factor.

For $T<T_{\mathrm{f}}(\sigma)$ the free barrier energy becomes independent of temperature equalling $F(T_{\mathrm{f}}(\sigma),\sigma)=E(T_{\mathrm{f}}(\sigma),\sigma)$. When sampling the $N'$ heterogeneous volumes of the RVE, the kinetic freezing temperature will fluctuate around this value due to fluctuations in the corresponding kinetic freezing barrier energy, the statistics of which is set by the extreme value Weibull distribution~\cite{Gumbel2004}. In this temperature regime, a heterogeneous volume element may contain no accessible barriers below a particular energy threshold allowing for the possibility that no thermally activated structural transformation occurs. It is only in this temperature regime that a single structural transformation associated with a single barrier energy is formally valid. This scenario is quite different from that of the statistics of the most probable, occurring at higher temperatures, where on average there exists a large number of accessible barrier energies resulting in a statistically meaningful average plastic transition rate for each heterogeneous volume. For Viteroly-1, the parameters of sec.~\ref{SecExp} give $T_{\mathrm{f}}\simeq191$ K indicating that below room temperature this strongly heterogeneous regime of statistics begins to dominate. When using the numerical values of the model parameters derived from a fit to the peak stress, the freezing temperature rises to approximately 203 K, which is somewhat smaller than that obtained when using a Gaussian distribution fitted also to the peak stress (as done in paper I).

\begin{figure}
\begin{center}
\includegraphics[clip,width=0.9\textwidth]{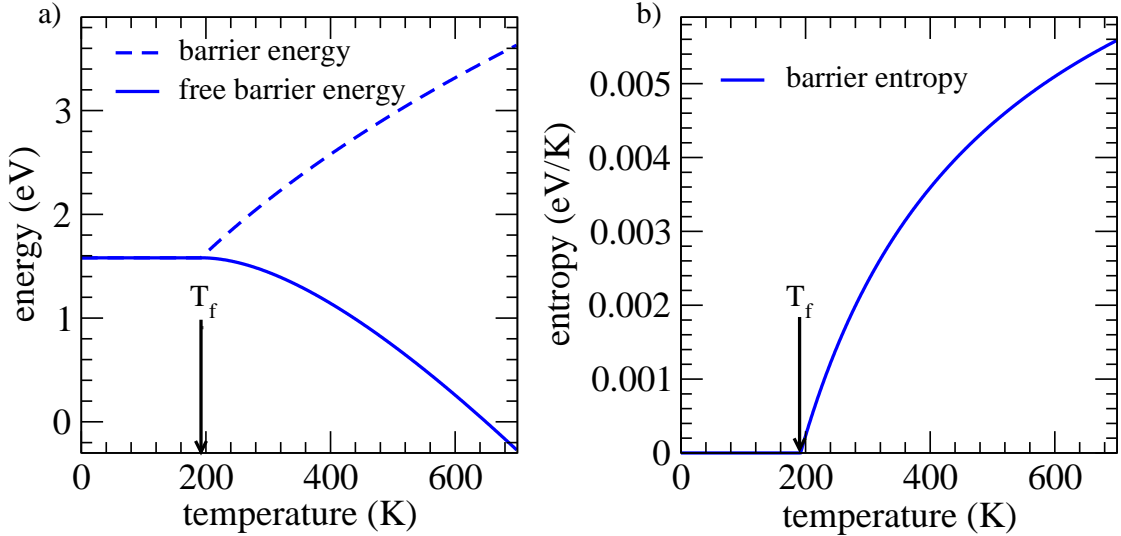}
\end{center}
\caption{a) Plot of free barrier energy and (apparent) barrier energy, and b) barrier entropy as a function of temperature for the Viteroly-1 parametrisation. Indicated is the freezing temperature, $T_{\mathrm{f}}$, below which the barrier entropy is zero and on average there exists one available structural transformation within each heterogeneous volume, resulting in the free energy and barrier energy becoming temperature independent quantities for $T<T_{\mathrm{f}}$.} \label{FigFreeEnergy} 
\end{figure}

Fig.~\ref{FigFreeEnergy} plots the temperature dependence of the free barrier energy, apparent barrier energy and barrier entropy for the flow stress parameters of Viteroly-1. Both the apparent barrier entropy and barrier entropy reduce with temperature, until the latter becomes zero at the freezing temperature, $T_{\mathrm{f}}$. Below this temperature range both quantities become constants independent of temperature. The corresponding free barrier energy shown in fig.~\ref{FigFreeEnergy}a is negative above a temperature regime associated with the glass transition, and rises when the temperature is lowered. According to eqn.~\ref{EqnFE}, in the temperature regime where the free barrier energy is positive, the plastic rate reduces many orders of magnitude when compared to the value associated with the glass transition temperature regime. At $T<T_{\mathrm{f}}$ the free barrier energy also becomes constant, and it is in this regime a simple temperature independent value for both the number of available structural and the energy barrier is formally justified. Despite this, the work of sec.~\ref{SecExp} demonstrates that temperature independent quantities can be used to describe reasonably well the high temperature/homogeneous deformation regime. Indeed fig.~\ref{FigFreeEnergy} demonstrates that both barrier energy and entropy do not vary greatly over the 550 K to 650 K temperature range.

The mean-field form for the shear strain rate, eqn.~\ref{EqnAvePSR4}, is in practice quite similar to the well known Arrhenius form, eqn.~\ref{EqnSimpleT}. Indeed it is operationally similar justifying the so-called critical barrier energy alluded to by Argon~\cite{Argon1979} and Johnson and Samwer~\cite{Johnson2005}. Indeed, in paper I, the fluctuations around the apparent barrier energy where shown to scale as $1/\sqrt{N}$. Despite this similarity, there exist some fundamental differences between the models. The Arrhenius shear strain rate is assumed to be proportional to the fraction of volume in which  structural transformations can take place~\cite{Schuh2007}. The present theory and eqn.~\ref{EqnAvePSR4} also gives a plastic shear rate that is proportional to the number of available structural transitions, $\exp(S(T)/k_{\mathrm{b}})$, however this number scales exponentially with atom number and therefore volume. Since this number can be rather large, the traditional athermal concept of a dilute density of (liquid like) regions existing within the structural glass which are amenable to plastic deformation has been abandoned --- in the present work, any particular region of the material can admit a very large number and variety of structural transformations. Of course, at low enough $T$ the apparent number of structural transformations drastically reduces as the entire $\alpha$-relaxation PEL gradually freezes out and the kinetic freezing regime is entered.

Another similarity between eqns.~\ref{EqnAvePSR4} and \ref{EqnSimpleT} is a shear stress dependent term which for the present work has the form $x\exp(x^{2})(\mathrm{I}_{0}[x^{2}]+\mathrm{I}_{1}[x^{2}])$ and for the Arrhenius form is $\sinh(x)$. In both cases $x=\mathrm{Volume}\times\sigma/(k_{\mathrm{b}}T)$. For the Arrhenius form, the volume term is seen as an activation volume associated with the barrier enthalpy of the volume distortion needed for the structural transformation to occur. In the present work, this quantity originates from the variance of the distribution of available local plastic strains, $\delta\gamma$, via $\delta\Omega=V_{0}\delta\gamma/2$. The right-hand-side of this equality may be seen as a measure of the variance of the available slipped volume within the system, which itself is coming from a characteristic internal slipped area multiplied by a characteristic slip distance.

In eqn.~\ref{EqnSimpleT}, the $\sinh(x)$ factor arises from reversibility of the single characteristic structural transformation of the system and that the sign of the enthalpy term changes for the inverse process. For the present work, the factor $x\exp(x^{2})(\mathrm{I}_{0}[x^{2}]+\mathrm{I}_{1}[x^{2}])$ in eqn.~\ref{EqnAvePSR4} is due to a quite different reason where each transition to a new mega-basin results in the material on average having once again access to all possible plastic transitions. Thus although the reverse transition is allowed, it will occur with negligible probability, since it must compete with all other accessible $\alpha$ structural transformations. The factor $x\exp(x^{2})(\mathrm{I}_{0}[x^{2}]+\mathrm{I}_{1}[x^{2}])$ then arises from the mean-field average (integration) over all possible plastic transitions.

Whilst the currently developed mean-field picture affords some insight into the consequences of the present theory to macro-plasticity, it certainly is limited to the low strain rate and high temperature range, a regime that is experimentally insensitive to fluctuations away from homogeneity, and also to the thermal and loading history of the material. Indeed, inherent to the above picture is that the material has no memory of its past state. This is direct result of the mean field theory variant presently used since the elastic interaction between heterogeneous volumes is not at all considered. When such interactions are taken into account the pre-history of the material will play an integral role in its response to an external condition. In this regard the internal stress field may be viewed as the state-variable of the model. To address these aspects of low temperature, high strain rate and material history, the mean field approach must be abandoned and an approach which considers correlated spatial variations in plastic activity must be considered. How to develop a theory of macroscopic plasticity beyond mean-field, and which spans the regime of $T_{\mathrm{f}}$ up to $T_{\mathrm{g}}$ for experimentally accessible strain rates, will be the subject of paper III in this series of work.

\section{Acknowledgements}

The authors wish to thank D. Rodney and K. Samwer for helpful discussions.

\end{document}